# Mobile-friendly Image de-noising: Hardware Conscious Optimization for Edge Application

Srinivas Miriyala*  
*On-Device AI*  
*Samsung Research Institute*  
Bangalore, India  
srinivas.m1@samsung.com

Sowmya Vajrala*  
*On-Device AI*  
*Samsung Research Institute*  
Bangalore, India  
v.lahari@samsung.com

Hitesh Kumar  
*On-Device AI*  
*Samsung Research Institute*  
Bangalore, India  
h1.kumar@samsung.com

Sravanth Kodavanti  
*On-Device AI*  
*Samsung Research Institute*  
Bangalore, India  
k.sravanth@samsung.com

Vikram Rajendiran  
*On-Device AI*  
*Samsung Research Institute*  
Bangalore, India  
vikram.nr@samsung.com

*Abstract*— Image enhancement is a critical task in computer vision and photography that is often entangled with noise. This renders the traditional Image Signal Processing (ISP) ineffective compared to the advances in deep learning. However, the success of such methods is increasingly associated with the ease of their deployment on edge devices, such as smartphones. This work presents a novel mobile-friendly network for image de-noising obtained with Entropy-Regularized differentiable Neural Architecture Search (NAS) on a hardware-aware search space for a U-Net architecture, which is first-of-its-kind. The designed model has 12% less parameters, with ~2-fold improvement in on-device latency and 1.5-fold improvement in the memory footprint for a 0.7% drop in PSNR, when deployed and profiled on Samsung Galaxy S24 Ultra. Compared to the SOTA Swin-Transformer for Image Restoration, the proposed network had competitive accuracy with ~18-fold reduction in GMACs. Further, the network was tested successfully for Gaussian de-noising with 3 intensities on 4 benchmarks and real-world de-noising on 1 benchmark demonstrating its generalization ability.

*Keywords—De-Noising, Differentiable NAS, Hardware aware Search space, Smartphone Deployment*

## I. Introduction

One critical application of Image Signal Processing (ISP) is De-noising [1, 2, 3, 4], where the goal is to improve the quality and visibility of images corrupted with noise artefact. Traditional methods often struggle to produce satisfactory results in such settings, prompting the exploration with deep learning. However, as deep learning continues to evolve, its migration towards edge necessitates the design of networks that are accurate and efficient for deployment on embedded devices. While one-shot super-net based strategies [5] remain the most effective methods for the efficient design of deep neural networks, they have been less explored for Image De-noising in the literature. In summary, the reported strategies for differentiable NAS face the following issues when scaled for the task of Image De-noising:

1. *Limited Search Space exploration.* Differentiable NAS works efficiently in simpler search spaces but struggle in complex domains such as Image restoration.

2. *Performance Collapse.* To facilitate training, the discrete optimization is smoothened, resulting in a bias for suboptimal parameter efficient networks..

3. *Indecisiveness* due to equal weightage on different subnets because of improper tuning of hyper-parameters (e.g. the temperature in Softmax).

4. *Optimization instabilities* arising from different scales of architecture parameters and weights in the networks.

5. *Issues in training* the massive super-net and inability to scale on resource-constrained environments.

The authors address these issues in the current work, where an Entropy Regularized Differentiable Search strategy is applied on a hardware aware search space (see Fig.1) to design mobile-friendly network for deployment on smartphones. The obtained model (Entropy Regularized Neural Network or **ERN-Net**) resulted in 2-fold improvement in latency when deployed and profiled on the Qualcomm's NPU on the Samsung Galaxy S24 (GS24) Ultra device with minimal loss in accuracy. The proposed approach is generic and can be used for deployment on any standard mobile device. The contributions of the proposed work are summarized below.

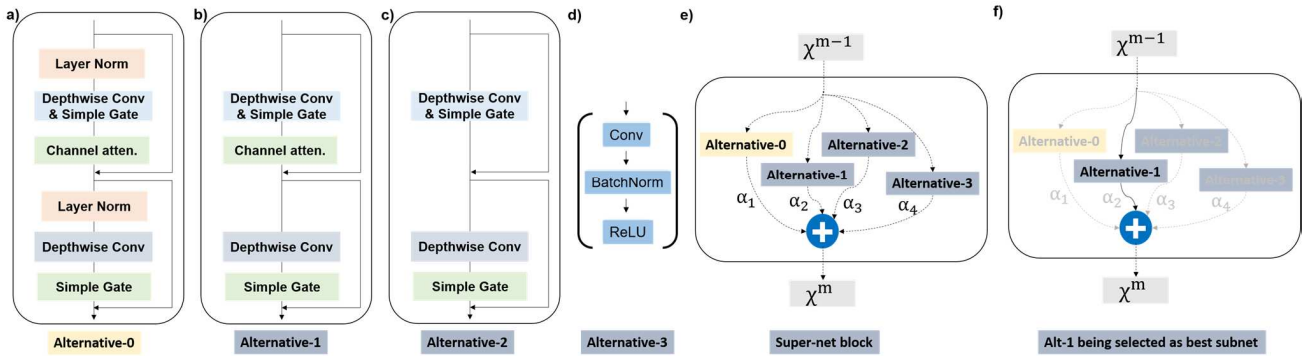

Figure 1. A pictorial representation of hardware aware search space with the a) base block, b) to d) hardware aware alternatives and e) the super-net block

*Equal Contribution

1. Differentiable NAS with hardware-aware search space on U-Net architectures for Image De-noising.
2. Normalized Entropy Regularization to prevent indecisiveness and improve convergence by scaling
3. Hardware conscious loss function design for performance enhancement with emphasis on accuracy
4. Emphasis on inference optimization of existing SOTA AI model for edge deployment.
5. Generalization to both synthetic and real-word noise in images and fine-tuning to new use-cases.
6. Edge deployment, profiling and comprehensive comparison of on-device performance indicators

## II. Proposed Method

### A. Hardware-aware Search Space

The search space for the Image De-noising was designed based on the 36-block NAFNet model [6]. NAFNet is a U-Net styled architecture with 4 encoders, 1 middle block and 4 decoders. Each of the encoder, middle block, and decoder consists of one or more Nonlinear Activation Free (NAF) block. The base network has a configuration of (2-2-4-8)-12-(2-2-2-2), where the numeric indicates the number of NAF blocks in each of (encoders)-middle-(decoders) blocks. Each NAF block comprises of Layer Normalization, depth-wise convolutions, channel attention and a gated approximation for GeLU activation as shown in Fig. 1a [6]. Given the hardware for deployment (e.g. the Neural Processing Unit on the Qualcomm's sm8650 chipset) the NAF block is profiled to identify the time-intensive operations and the following inferences are drawn.

1. Since the Layer Normalization (LN) operation involves the calculation of normalization parameters during the runtime, it consumes the maximum time.
2. Since the channel attention mechanism involves global average pooling, which necessitates the matrix operation as a whole, it was found to be computationally expensive after the LN operation.

Table 1. Hardware aware Search Space & the weight distribution of candidates obtained after applying Entropy Regularized Differentiable NAS for de-noising SIDD. The base network configuration is (2-2-4-8)-12-(2-2-2-2). The middle block with 12 NAF blocks is not included in the search space as its latency is negligible due to the very low resolution processed by it. Fig.1 illustrates the architecture of the considered Alternatives.

| Alternatives (see Fig. 1) | Encoder 1 | Encoder 2 | Encoder 3 | Encoder 4 | Decoder 4 | Decoder 3 | Decoder 2 | Decoder 1 |
|---|---|---|---|---|---|---|---|---|
| 1 x Alternative-3 | 3.1E-09 | **9.9E-01** | 1.2E-05 | 7.9E-10 | **9.9E-01** | 6.2E-08 | 1.2E-07 | 7.7E-09 |
| 2 x Alternative-3 | 3.1E-09 | 2.4E-07 | **9.9E-01** | 4.6E-04 | 2.1E-05 | **9.9E-01** | **9.9E-01** | 7.8E-09 |
| 3 x Alternative-3 | 3.1E-09 | 4.6E-06 | 1.3E-06 | 1.4E-07 | 2.4E-08 | 1.1E-05 | 7.7E-06 | 7.7E-09 |
| 4 x Alternative-3 | 3.3E-07 | 7.2E-08 | 1.4E-07 | 2.1E-08 | 4.3E-07 | 1.1E-06 | 7.5E-06 | 4.4E-06 |
| 5 x Alternative-3 | NA | NA | NA | 9.8E-09 | NA | NA | NA | NA |
| 6 x Alternative-3 | NA | NA | NA | 7.9E-10 | NA | NA | NA | NA |
| 7 x Alternative-3 | NA | NA | NA | 6.4E-07 | NA | NA | NA | NA |
| 8 x Alternative-3 | NA | NA | NA | 8.7E-06 | NA | NA | NA | NA |
| 1 x Alternative-0 | 9.1E-06 | 1.3E-05 | 1.5E-06 | 5.1E-09 | 1.2E-09 | 9.5E-06 | 2.4E-05 | **9.9E-01** |
| 2 x Alternative-0 | **9.9E-01** | 1.1E-05 | 4.4E-06 | 8.1E-10 | 1.7E-07 | 1.1E-05 | 7.3E-05 | 5.8E-06 |
| 3 x Alternative-0 | NA | NA | 2.9E-06 | 7.9E-10 | NA | NA | NA | NA |
| 4 x Alternative-0 | NA | NA | 4.1E-06 | 8.1E-10 | NA | NA | NA | NA |
| 5 x Alternative-0 | NA | NA | NA | 8.1E-10 | NA | NA | NA | NA |
| 6 x Alternative-0 | NA | NA | NA | 7.9E-10 | NA | NA | NA | NA |
| 7 x Alternative-0 | NA | NA | NA | 8.1E-10 | NA | NA | NA | NA |
| 8 x Alternative-0 | NA | NA | NA | 7.9E-10 | NA | NA | NA | NA |
| 1 x Alternative-1 | 1.1E-08 | 5.7E-09 | 3.1E-09 | 8.1E-10 | 1.1E-09 | 4.4E-09 | 8.5E-09 | 1.7E-07 |
| 2 x Alternative-1 | 2.6E-08 | 5.8E-09 | 3.1E-09 | 8.1E-10 | 1.6E-08 | 4.4E-09 | 8.6E-09 | 5.8E-08 |
| 3 x Alternative-1 | NA | NA | 3.1E-09 | 2.5E-09 | NA | NA | NA | NA |
| 4 x Alternative-1 | NA | NA | 3.1E-09 | 8.0E-10 | NA | NA | NA | NA |
| 5 x Alternative-1 | NA | NA | NA | 3.4E-09 | NA | NA | NA | NA |
| 6 x Alternative-1 | NA | NA | NA | 8.1E-10 | NA | NA | NA | NA |
| 7 x Alternative-1 | NA | NA | NA | 8.1E-10 | NA | NA | NA | NA |
| 8 x Alternative-1 | NA | NA | NA | 8.1E-10 | NA | NA | NA | NA |
| 1 x Alternative-2 | 3.2E-09 | 5.9E-09 | 3.2E-06 | 7.9E-10 | 3.7E-07 | 6.8E-06 | 3.2E-05 | 3.2E-06 |
| 2 x Alternative-2 | 9.1E-06 | 1.8E-05 | 9.1E-07 | 4.5E-08 | 2.8E-07 | 7.2E-06 | 3.1E-05 | 1.5E-05 |
| 3 x Alternative-2 | NA | NA | 1.2E-06 | 2.9E-02 | NA | NA | NA | NA |
| 4 x Alternative-2 | NA | NA | 2.1E-06 | 7.9E-10 | NA | NA | NA | NA |
| 5 x Alternative-2 | NA | NA | NA | 2.1E-08 | NA | NA | NA | NA |
| 6 x Alternative-2 | NA | NA | NA | **4.8E-01** | NA | NA | NA | NA |
| 7 x Alternative-2 | NA | NA | NA | **4.8E-01** | NA | NA | NA | NA |
| 8 x Alternative-2 | NA | NA | NA | 8.5E-08 | NA | NA | NA | NA |

3. Further, the necessity of NAF blocks in every encoder or decoder is challenged by attempting to replace with the simple {Convolution - Batch Normalization – ReLU} blocks along with Residual Connection. During compilation, such a block can be folded into a single operation eliminating the need for intermediate data transfer between the processor and memory, thus reducing the latency drastically.
4. Finally, the number of NAF blocks in each encoder/middle/decoder block can be minimized.

Using these observations from the on-device profiling, 3 different alternatives to the NAF block are prepared as presented in Fig. 1b to 1d. Further, to confirm the inferences, these alternatives are profiled on the device and compared with the NAF block to confirm they have lower GFLOPs and latency, thus making the search space mobile-friendly. The weighted combination of all the alternatives forms the super-net (see Fig. 1e) from which Differentiable NAS can determine the optimal combination with highest accuracy (see Fig. 1f).

### B. Entropy Regularized Differentiable NAS

The conventional formulation for differentiable NAS is shown from Eq. 1 to 10. The output map $\chi^m$ from a single $m^{th}$ block (either encoder or decoder or middle block) is evaluated as the weighted sum of outputs from each of the K alternatives as shown in Eq. 1. Here, $f_i(\chi^{m-1})$ is the output of the $i^{th}$ alternative to the $m^{th}$ block. The weights $\alpha_i \forall i = 1$ to K are discrete probabilities generated from the architectural parameters $\varphi_i \forall i = 1$ to K on each alternative as shown in Eq. 2. These weights ($\alpha$) are called the architectural encodings.

$$\chi^m = \sum_{i=1}^{K} \alpha_i [f_i(\chi^{m-1})] \tag{1}$$

$$\alpha_i = \text{Softmax}(\varphi_i \forall i = 1 \text{ to K}) \tag{2}$$

This is repeated on all the blocks of the network to generate output map $\chi^{out}$, which is compared with the Ground Truth $\chi^{GT}$ to generate the training loss $L_T$ as shown in Eq. 3 where $\mathcal{F}$ is a suitable Loss function such as $L_2$ loss. During training of the super-net, $L_T$ is minimized with respect to $\{\varphi_i \forall i = 1 \text{ to K}\}_{\text{for all blocks}}$ and all the weights $w$ in the network.

$$L_T \leftarrow \mathcal{F}(\chi^{out}, \chi^{GT}) \tag{3}$$

To ensure the optimizer selects the most efficient alternative and avoid the base NAF block, each alternative is associated with a penalty $\wp_i \forall i = 1$ to K that correlates with the on-device latency and GFLOPs of the alternatives. These penalties are then weighted by the architectural encodings. The weighted sum of penalties accumulated over all the M blocks $L_P$ as shown in Eq. 4 is added to the training loss that will be minimized during training the super-net. These aspects resolve the issues of limited search space and performance collapse.

$$L_P = \sum_{m=1}^{M} \left[ \sum_{i=1}^{K} \alpha_i [\wp_i] \right]_m \tag{4}$$

$$L_{ER} = -\sum_{m=1}^{M} \left[ \sum_{i=1}^{K} \alpha_i \log(\alpha_i) \right]_m \tag{5}$$

The issues of indecisiveness and optimization instabilities are handled with the normalized Entropy Regularization (ER). In this approach, an entropy penalty on the Softmax distribution of architectural parameters is introduced to discourage equal weighting across the alternatives (see Eq. 5). The entropy term is added to the loss function, pushing the Softmax to prefer more decisive selections by reducing the uncertainty in the distribution. Further, the strength of this regularization is scaled by the parameter $\lambda$ that is scheduled exponentially with the epochs, introducing the normalizing effect to improve the convergence. The ER term is shown in Eq. 5 while the overall loss L for training the super-net is shown in Eq. 6.

Table 2. Comparison of accuracy and complexity among various SOTAs for raw image de-noising. The proposed network is called ERN-Net

| Dataset | | MIRNet [15] | MAXIM [16] | Restormer [17] | RIDNet [18] | Uformer [19] | KBNet [20] | NAFNet [6] | ERN-Net |
|---|---|---|---|---|---|---|---|---|---|
| SIDD | PSNR | 39.72 | 39.89 | 40.02 | 38.71 | 39.75 | 40.35 | 43.42 | 43.09 |
| | SSIM | 0.959 | 0.960 | 0.960 | 0.914 | 0.959 | 0.972 | 0.959 | 0.956 |
| | GMACs | 786 | 169.5 | 140 | 89 | 88.8 | 57.8 | 65 | 42 |
| Sense Noise | PSNR | 35.30 | 35.43 | 35.52 | 34.88 | 35.43 | 35.6 | 35.55 | 34.6 |
| | SSIM | 0.919 | 0.922 | 0.924 | 0.915 | 0.920 | 0.924 | 0.923 | 0.988 |
| | GMACs | 130 | 120 | 80 | 89 | 90 | 57.8 | 65 | 42 |

Table 3. Comparison of PSNR among various SOTAs for removing Gaussian noise of different standard deviations ($\sigma$)

| Datasets | McMaster [9] | | | Urban100 [10] | | | Kodak 24 [11] | | | CBSD 68 [12] | | | MACs |
|---|---|---|---|---|---|---|---|---|---|---|---|---|---|
| Model | $\sigma = 15$ | $\sigma = 25$ | $\sigma = 50$ | $\sigma = 15$ | $\sigma = 25$ | $\sigma = 50$ | $\sigma = 15$ | $\sigma = 25$ | $\sigma = 50$ | $\sigma = 15$ | $\sigma = 25$ | $\sigma = 50$ | |
| RDN [21] | - | - | - | - | - | 29.38 | - | - | 29.66 | - | - | 28.31 | 1.4T |
| SwinIR [13] | 35.61 | 33.20 | 30.22 | 35.13 | 32.90 | 29.82 | 35.34 | 32.89 | 29.79 | 34.42 | 31.78 | 28.56 | 759G |
| IPT [22] | - | - | 29.98 | - | - | 29.71 | - | - | 29.64 | - | - | 28.39 | 512G |
| RNAN [23] | - | - | 29.72 | - | 31.99 | 29.08 | - | - | 29.58 | - | - | 28.27 | 496G |
| DRUNet [24] | 35.40 | 33.14 | 30.08 | 34.81 | - | 29.61 | 35.31 | 32.89 | 29.86 | 34.30 | 31.69 | 28.51 | 144G |
| Restormer [17] | 35.61 | 33.34 | 30.30 | 35.13 | 32.96 | 30.02 | 35.47 | 33.04 | 30.01 | 34.40 | 31.79 | 28.60 | 141G |
| KBNet [20] | 35.56 | 33.31 | 30.27 | 35.15 | 32.96 | 30.04 | 35.46 | 33.05 | 30.04 | 34.41 | 31.80 | 28.62 | 69G |
| DnCNN [25] | 33.45 | 31.52 | 28.62 | 32.98 | 31.40 | 27.59 | 34.60 | 32.14 | 28.95 | 33.90 | 31.24 | 27.95 | 37G |
| NAFNet [6] | 28.61 | 25.92 | 19.67 | 26.28 | 23.96 | 17.95 | 29.19 | 26.95 | 20.34 | 28.58 | 26.14 | 18.54 | 65G |
| ERN-Net (ours) | 28.47 | 26.86 | 23.28 | 24.95 | 23.79 | 21.21 | 28.62 | 27.01 | 23.42 | 27.88 | 26.27 | 22.89 | 42G |

$$\{\boldsymbol{w}, \boldsymbol{\varphi}\}^* = \mathrm{argmin}(L = L_T + L_P + L_{ER}) \quad (6)$$

Following the convention of differentiable NAS [7], upon training, the alternative with highest encoding is selected for each block and fine-tuned to achieve the desired accuracy.

### III. RESULTS & DISCUSSIONS

#### A. Raw Image De-noising

The super-net is trained to de-noise the SIDD dataset [8]. The weight distribution obtained upon the convergence of training for each block in the network is shown in Table 1. The alternatives with highest weights (highlighted in Table 1) replace the original blocks in the base network. The NAS-optimized network is fine-tuned further to obtain the desired convergence in accuracy. When profiled on GS24 ultra, it had a latency of 110ms, while the base NAFNet model's inference required 220ms. The obtained model resulted in 12% less parameters, ~2 fold improvement in on-device latency and 13% improvement in memory with 0.7% and 0.3% drop in PSNR and SSIM, respectively.

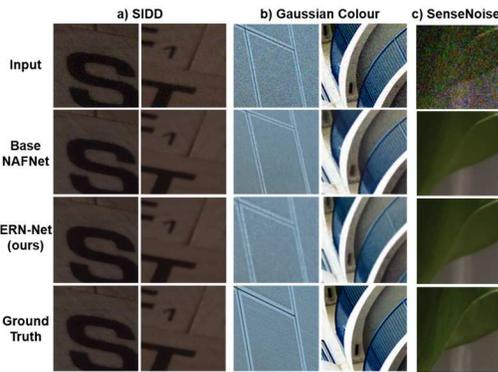

Figure 2. Visual comparison of outputs from ERN-Net and the base model for de-noising Gaussian and raw noise (SIDD & SenseNoise)

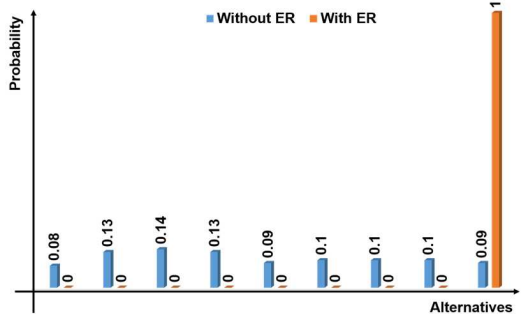

Figure 3. Ablation analysis to check the effect of Entropy Regularization on last decoder of the network with 9 different alternatives

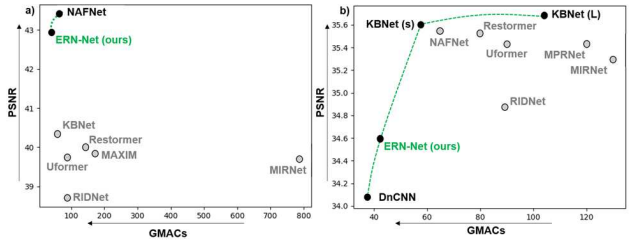

Figure 4. The objective function space showing the trade-off between PSNR (higher the better) and GMACs (lower the better) for a) SIDD and b) SenseNoise dataset. The green dashed curve is the non-dominated front.

The performance of the network is compared with SOTA models for filtering the raw noise in SIDD dataset in Table 2. It is important to note that all the models that are being compared here (including base model and NAS optimized model) are trained on SIDD [8]. The visualization results for SIDD is presented in Fig. 2a.

#### B. Filtering Gaussian Noise

To test the generalization ability of the obtained network, it was deployed to filter the Gaussian noise of varying intensities on color images of four different datasets, namely McMaster [9], Urban 100 [10], Kodak 24 [11], and CBSD68 [12]. The results are compared with SOTA networks in Table 3. While the NAS model outperforms the base model, when compared with the SOTA Swin Transformer for Image Restoration [13], competitive Image Quality is obtained with a reduction of ~18 folds in GMACs. The inputs, ground truth and outputs of various models can be visualized in Fig. 2b.

#### C. Generalization on test dataset

The obtained model was also tested for filtering the raw noise on another open source dataset called SenseNoise [14] that serves as test data for checking the capabilities of the network trained on SIDD. The visualization results are presented in Fig. 2c. The comparison of model performance (profiled on GS24) and generalization on SenseNoise is presented in Table 2. Compared with MIRNet [15] for the case of SenseNoise, the NAS model is able to generate the best PSNR values with a reduction of 18 folds in GMACs, indicating the prowess of proposed method for optimization and commercialization of complex AI networks.

The effect of entropy regularization is visualized in Fig. 3 when differentiable search was carried out on only the last decoder of the network with 9 different alternatives. The Pareto trade-off between PSNR and MACs in case of raw image de-noising for SIDD and SenseNoise datasets can be visualized in Figs. 4a and b, respectively. These figures compare various SOTAs where it is observed that in both cases, the NAS-optimized model forms part of the non-dominated Pareto front, demonstrating the prowess of the proposed method.

### IV. CONCLUSION

This work presents an effort to optimize pre-trained foundation models in Image De-noising for deployment on smartphones. The proposed framework offers network surgery, hardware-aware search space design and a novel Entropy regularised differentiable search strategy that is first-of-its-kind. The tasks of development and deployment of complex AI models for image de-noising are both resolved successfully without intermixing them or downplaying one on the grounds of other. Reformulating NAS as a strategy to optimize the SOTA can thus, capitalize on the already built models and ensure their full potential by deploying on embedded devices. As part of the future scope, such a generic strategy can be extensively applied for optimization and deployment of compute and memory-intensive models for Generative Artificial Intelligence.

## REFERENCES

[1] Vincent, P., Larochelle, H., Lajoie, I., Bengio, Y., & Manzagol, P. A. (2008). Stacked Denoising Autoencoders: Learning Useful Representations in a Deep Network with a Local Denoising Criterion. Journal of Machine Learning Research.

[2] Zhang, K., Zuo, W., Chen, Y., Meng, D., & Zhang, L. (2017). Beyond a Gaussian Denoiser: Residual Learning of Deep CNN for Image Denoising. IEEE Transactions on Image Processing.

[3] Lehtinen, J., Munkberg, J., Hasselgren, J., Laine, S., Karras, T., Aittala, M., & Aila, T. (2018). Noise2Noise: Learning Image Restoration without Clean Data. arXiv preprint arXiv:1803.04189.

[4] Izadi, S., Sutton, D., & Hamarneh, G. (2023). Image denoising in the deep learning era. Artificial Intelligence Review, 56(7), 5929-5974.

[5] He, X., Zhao, K., & Chu, X. (2021). AutoML: A survey of the state-of-the-art. *Knowledge-based systems*, *212*, 106622.

[6] Chen, L., Chu, X., Zhang, X., & Sun, J. Simple baselines for image restoration. In European conference on computer vision, 17-33 (2022).

[7] Liu, H., Simonyan, K., & Yang, Y. (2018). Darts: Differentiable architecture search. *arXiv preprint arXiv:1806.09055*.

[8] Abdelhamed, A., Lin, S., & Brown, M. S. A high-quality denoising dataset for smartphone cameras. In Proceedings of the IEEE conference on CVPR, 1692-1700 (2018).

[9] Zhang, L., Wu, X., Buades, A., & Li, X. (2011). Color demosaicking by local directional interpolation and nonlocal adaptive thresholding. *Journal of Electronic imaging*, *20*(2), 023016-023016.

[10] Huang, J. B., Singh, A., & Ahuja, N. (2015). Single image super-resolution from transformed self-exemplars. In *Proceedings of the IEEE conference on computer vision and pattern recognition* (pp. 5197-5206).

[11] Rich Franzen. Kodak lossless true color image suite. http://r0k.us/graphics/kodak/, 1999. Online accessed 24 Oct 2021

[12] Martin, D., Fowlkes, C., Tal, D., & Malik, J. (2001, July). A database of human segmented natural images and its application to evaluating segmentation algorithms and measuring ecological statistics. In *Proceedings eighth IEEE international conference on computer vision. ICCV 2001* (Vol. 2, pp. 416-423). IEEE.

[13] Liang, J., Cao, J., Sun, G., Zhang, K., Van Gool, L., & Timofte, R. (2021). Swinir: Image restoration using swin transformer. In *Proceedings of the IEEE/CVF international conference on computer vision* (pp. 1833-1844).

[14] Zhang, Y., Li, D., Law, K. L., Wang, X., Qin, H., & Li, H. (2022). Idr: Self-supervised image denoising via iterative data refinement. In *Proceedings of the IEEE/CVF conference on computer vision and pattern recognition* (pp. 2098-2107).

[15] Zamir, S. W., Arora, A., Khan, S., Hayat, M., Khan, F. S., Yang, M. H., & Shao, L. (2020). Learning enriched features for real image restoration and enhancement. In *Computer Vision–ECCV 2020: 16th European Conference, Glasgow, UK, August 23–28, 2020, Proceedings, Part XXV 16* (pp. 492-511). Springer International Publishing.

[16] Tu, Z., Talebi, H., Zhang, H., Yang, F., Milanfar, P., Bovik, A., & Li, Y. (2022). Maxim: Multi-axis mlp for image processing. In *Proceedings of the IEEE/CVF conference on computer vision and pattern recognition* (pp. 5769-5780).

[17] Zamir, S. W., Arora, A., Khan, S., Hayat, M., Khan, F. S., & Yang, M. H. (2022). Restormer: Efficient transformer for high-resolution image restoration. In *Proceedings of the IEEE/CVF conference on computer vision and pattern recognition* (pp. 5728-5739).

[18] Anwar, S., & Barnes, N. (2019). Real image denoising with feature attention. In *Proceedings of the IEEE/CVF international conference on computer vision* (pp. 3155-3164).

[19] Wang, Z., Cun, X., Bao, J., Zhou, W., Liu, J., & Li, H. (2022). Uformer: A general u-shaped transformer for image restoration. In *Proceedings of the IEEE/CVF conference on computer vision and pattern recognition* (pp. 17683-17693).

[20] Zhang, Y., Li, D., Shi, X., He, D., Song, K., Wang, X., ... & Li, H. (2023). Kbnet: Kernel basis network for image restoration. *arXiv preprint arXiv:2303.02881*.

[21] Zhang, Y., Tian, Y., Kong, Y., Zhong, B., & Fu, Y. (2018). Residual dense network for image super-resolution. In *Proceedings of the IEEE conference on computer vision and pattern recognition* (pp. 2472-2481).

[22] Chen, H., Wang, Y., Guo, T., Xu, C., Deng, Y., Liu, Z., ... & Gao, W. (2021). Pre-trained image processing transformer. In *Proceedings of the IEEE/CVF conference on computer vision and pattern recognition* (pp. 12299-12310).

[23] Zhang, Y., Li, K., Li, K., Zhong, B., & Fu, Y. (2019). Residual non-local attention networks for image restoration. *arXiv preprint arXiv:1903.10082*.

[24] Zhang, K., Li, Y., Zuo, W., Zhang, L., Van Gool, L., & Timofte, R. (2021). Plug-and-play image restoration with deep denoiser prior. *IEEE Transactions on Pattern Analysis and Machine Intelligence*, *44*(10), 6360-6376.

[25] Zhang, K., Zuo, W., Chen, Y., Meng, D., & Zhang, L. (2017). Beyond a gaussian denoiser: Residual learning of deep cnn for image denoising. *IEEE transactions on image processing*, *26*(7), 3142-3155.

.